\documentclass{JINST}

\title{Reactive Bonding Film for Bonding Carbon Foam Through Metal Extrusion}

\author{M. Chertok$^*$, M. Fu, M. Irving, C. Neher, M. Tripathi, R. Wang, G. Zheng\\
University of California, Davis\\
One Shields Ave., Davis, CA 95616, USA\\
\llap{$^*$} 
  E-mail: \email{chertok@physics.ucdavis.edu}\\
  }

\abstract{Future tracking detectors, such as those under development for the High Luminosity LHC, will require mechanical structures employing novel materials to reduce mass while providing excellent strength, thermal conductivity, and radiation tolerance.  Adhesion methods for such materials are under study at present.  This paper demonstrates the use of reactive bonding film as an adhesion method for bonding carbon foam.}

\begin{document}

\section{Introduction}
The CMS \cite{cmspaper,cmstracker,cmspix} and ATLAS \cite{atlaspaper,atlasibl} experiments at the CERN Large Hadron Collider are currently in the R\&D phase to prepare detector upgrades for the High-Luminosity LHC era, with a decade-long physics program starting around 2026.  Next generation silicon tracking detectors, for these and other experiments, will employ novel materials to reduce mass while meeting stringent requirements for thermal conductivity, stiffness, and radiation tolerance \cite{ibeam,lhcb,plume,snowmass, atlas-stave}.  A novel structural material for the upgrades will be carbon foam, which is a highly porous material (10\% dense), has a suitably low atomic number, and is tolerant to radiation exposure. Carbon foam also possesses high thermal conductivity ( > 20 W/m-K) and good strength-to-weight ratio \cite{snowmass,allcomp}.

Such upgraded detectors will require many components to be bonded together mechanically and thermally.  Bonding options include thermally conductive epoxy and thermally conductive tape. Each has its own advantages and difficulties, detailed in \cite{viehhauser, bonad-thesis, thermalpaper}.

Another option, reactive bonding film (RBF), is a multi-layer foil with alternating thin layers of aluminum and nickel \cite{indium}.  When activated with an electric spark or source of heat, an exothermic reaction occurs as these layers mix, generating localized temperatures of roughly 1500$^0$ C in less than 1 ms.  The produced heat can be used to instantly solder or weld materials together.  However, as this paper shows, it can also be used to extrude metals into porous media. Because the extremely fast and short duration heating keeps the heat localized during the process, surrounding components are not exposed to the high temperatures. The process is often used to connect silicon electronics to heatsinks, or to solder metals \cite{indium}. After the reaction, the RBF becomes nickel aluminide, which is electrically and thermally conductive.

In this paper, we explore RBF as a material for bonding together carbon foam structures.  By using aluminum foil as an in-between layer, the heat from the RBF extrudes the foil into the carbon foam forming "lock and key" shapes at the interface of the foam and foil.  These shapes hold the carbon foam layers together mechanically and allow for good thermal conduction properties.

\section{Method}
\subsection{Bonding Procedure}

\begin{figure}[!ht]
\centering
    \includegraphics[width=0.65\textwidth]{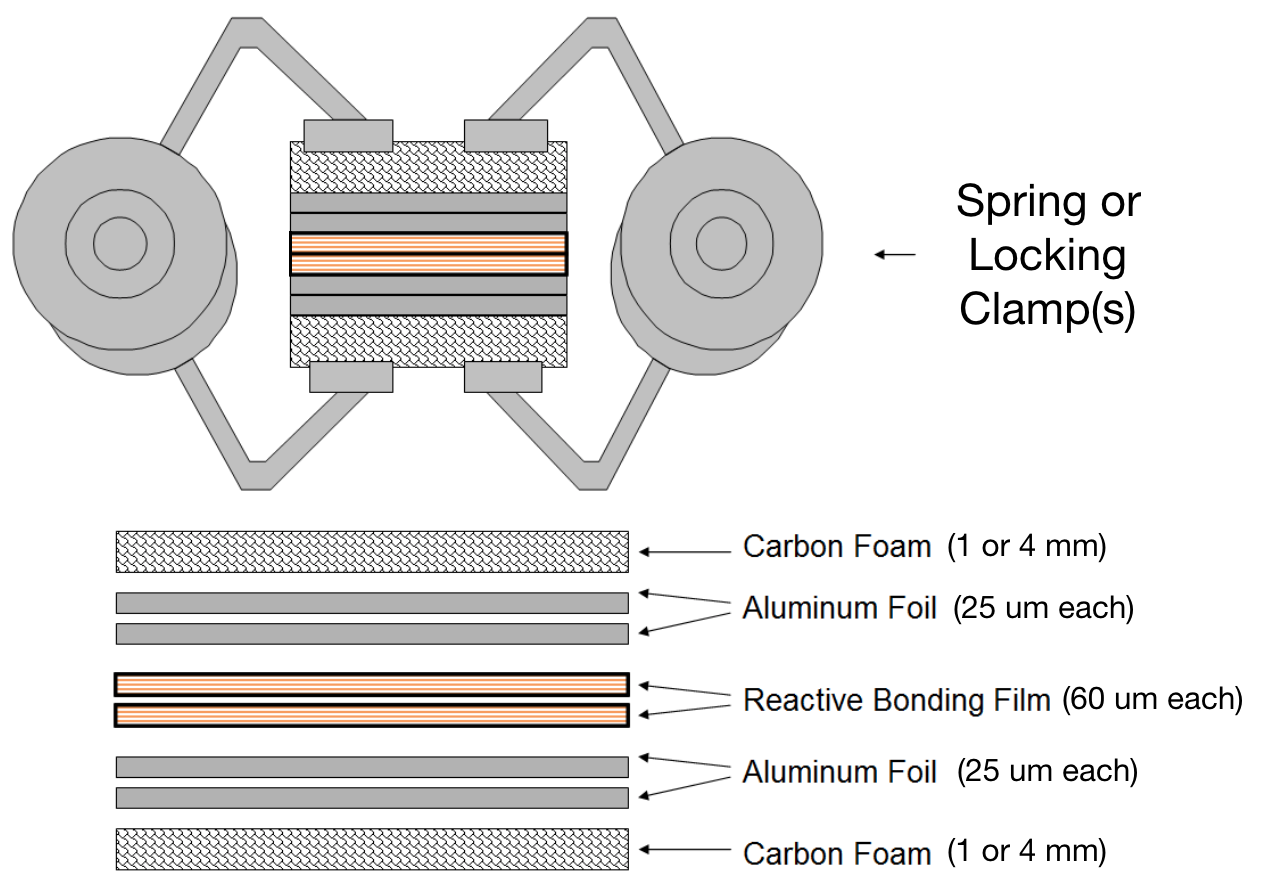}
  \caption{Upper: schematic of stack before ignition.  The stack is clamped together to allow the produced heat to bond the layers.  Lower: detailed view of stack, with example thicknesses and number of sheets of aluminum and RBF.  Not to scale.}
  \label{fig:stack}
\end{figure}

\begin{figure}[!ht]
\centering
       \includegraphics[width=0.6\textwidth]{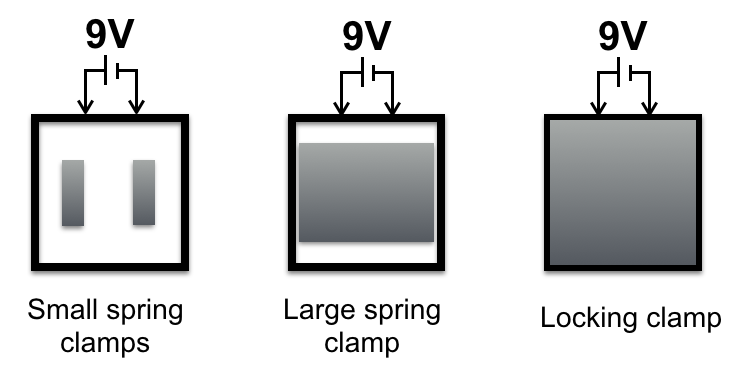}
  \caption{Top view of stack arrangements for the three clamping methods tested.  The shaded areas represent the clamping contacts, and the battery leads were touched to the RBF foil along one edge as shown.}
\label{fig:topview}
\end{figure}

Samples are composed of layers of carbon foam, aluminum foil, and RBF.  
The carbon foam is Allcomp "K9", a 10\% dense 130 pores per inch graphitized foam block \cite{allcomp}, which we section into sheets using a slitting saw.  SEM characterization of the foam surface after cutting gives an upper bound on roughness of 50 $\mu{\rm m}$.  The aluminum foil is 25 $\mu{\rm m}$ thick UHV grade foil of Aluminum 1235. The RBF is untinned 60 $\mu{\rm m}$ "Nanofoil" from Indium Corporation \cite{indium}.
Squares 2.5 cm on a side of each material are cut and sandwiched together to form a stack which has RBF in the center and carbon foam on the outsides, with  aluminum foil between each RBF - carbon foam interface, as shown in Figs.~\ref{fig:stack} and ~\ref{fig:topview}.  This stack is held together with one or two clamps while the RBF is ignited using a 9V battery, following the manufacturer's procedure \cite{indium}.  The sample is allowed to cool and released from the clamp(s) before being bonded to mounts for tensile strength testing.

The following variables were considered in the trials and adjusted accordingly to produce a robust and reproducible bond: number of aluminum foil sheets, number of RBF sheets, thickness of carbon foam, and clamping strength and uniformity.  Increasing the number of aluminum foil sheets increases the amount of material to be extruded into the carbon foam, and to a lesser extent, how much RBF is required to extrude the aluminum.  The number of RBF sheets determines how much heat is produced during the reaction.  The thickness of the carbon foam is varied to study possible effects on heat flow during the reaction. Finally, the clamping method affects how well the stack stays together during the reaction. Experimentation continued until a good combination of these experimental factors created a strong and uniform bond.

Various methods of clamping were investigated: a pair of small metal spring clamps, a large spring clamp with swiveling pads, and a mechanically locking clamp.  The small metal spring clamps provided 2 points of contact, with 12\% total coverage.  The large spring clamp covered 70\% of the surface of the samples.  The mechanically locking clamp fully covered the surface area of the samples.  Loading force and spring constants were measured using digital calipers and spring scales with clamps opened to the same position they would engage a sample made with 4 mm carbon foam. The small clamps engage with 20 N force and a 10 N/mm spring constant. The large clamps engage with 100 N force and a 20 N/mm spring constant. Finally, The locking clamps engage with 1 kN force and a 4 kN/mm spring constant.

\subsection{Testing procedure}
\begin{figure}[!ht]
\centering
      \includegraphics[width=0.35\textwidth]{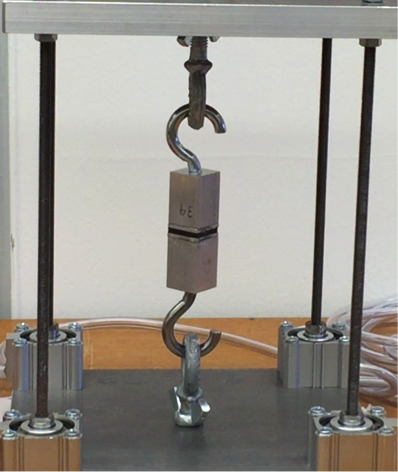}
    \includegraphics[width=0.6\textwidth]{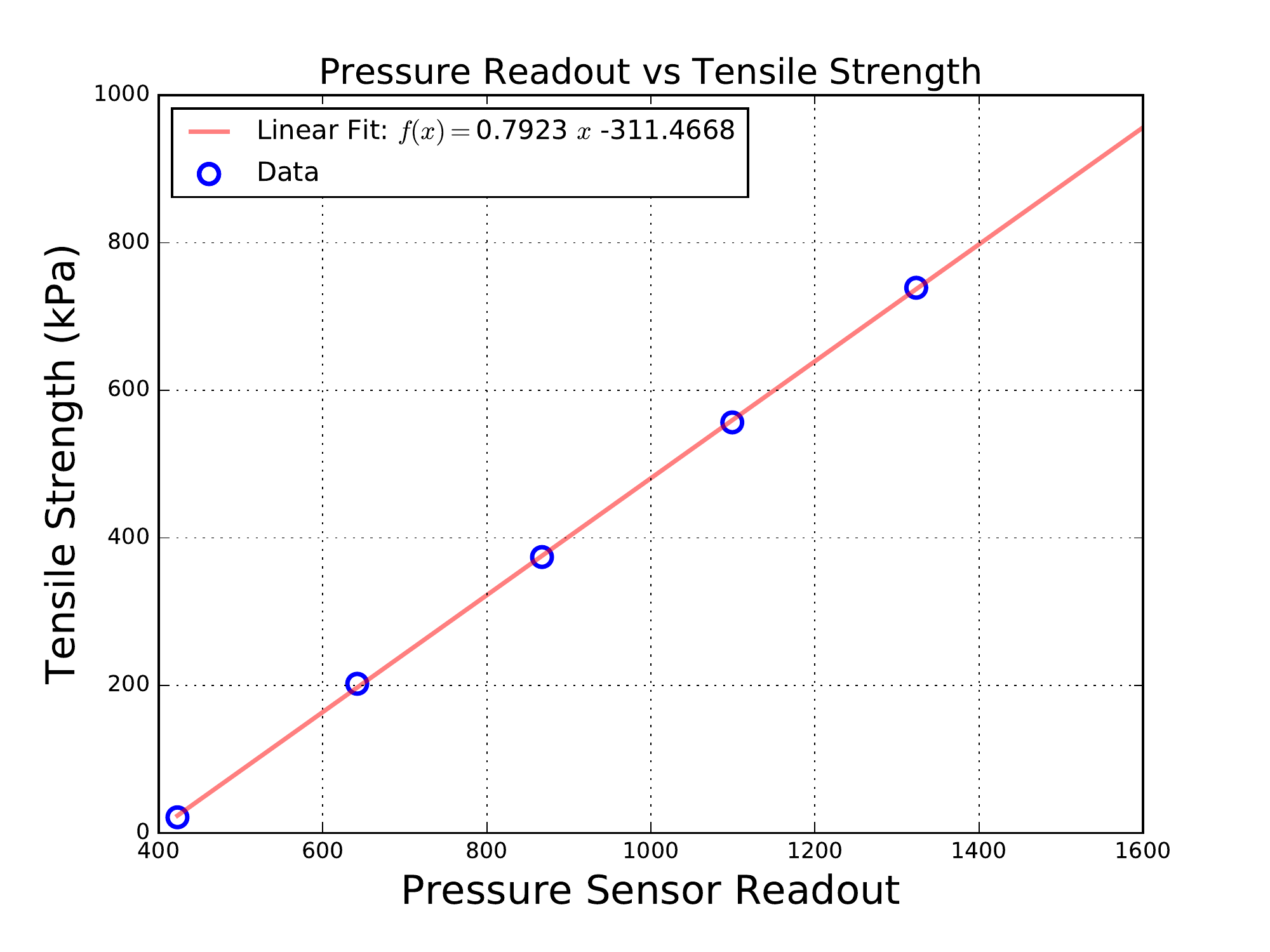}

  \caption{Left: Tensile strength testing apparatus with sample installed. Right: calibration results.}
  \label{fig:tensile}
\end{figure}

\begin{figure}[!ht]
\centering
    \includegraphics[width=0.5\textwidth]{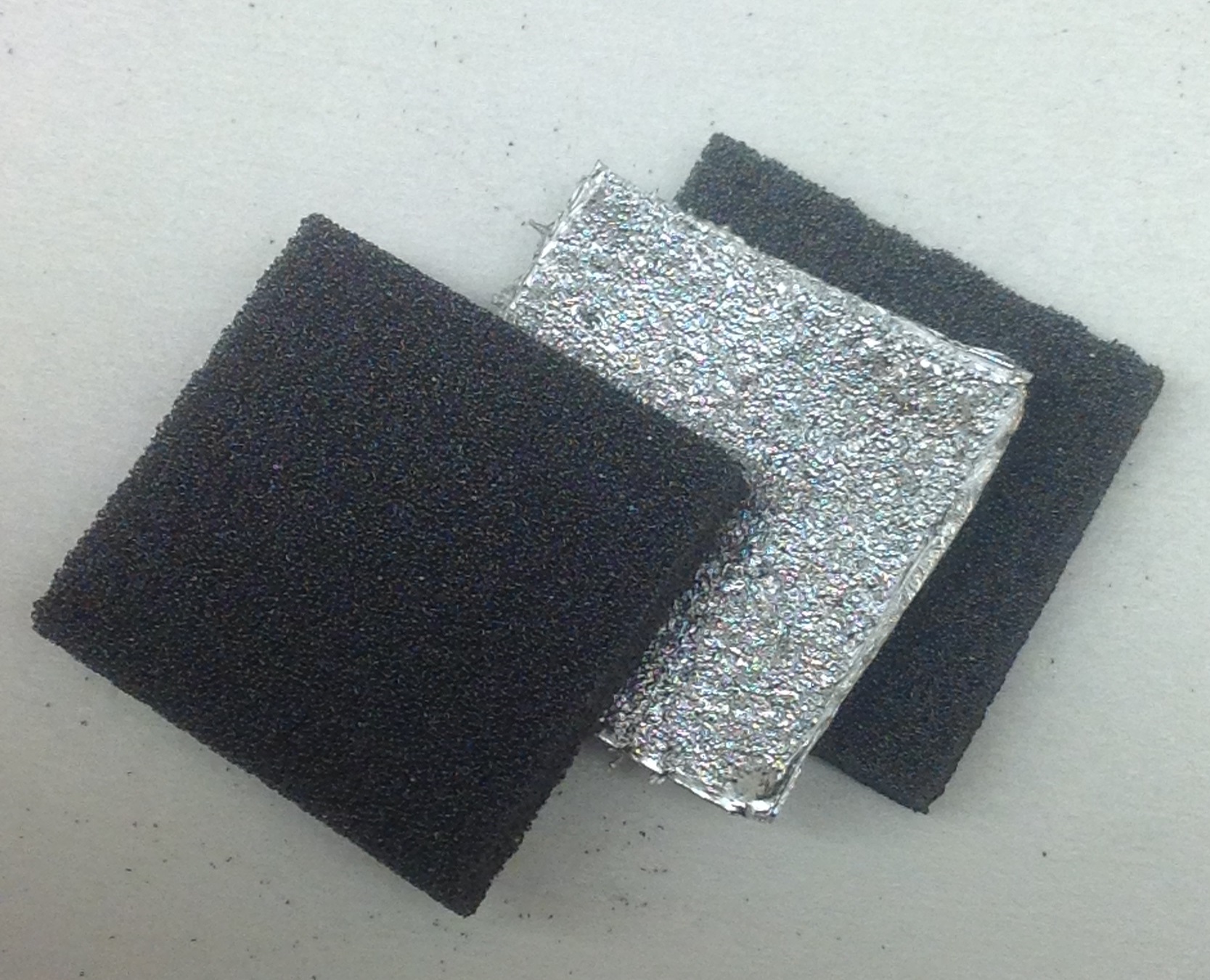}
  \caption{Mechanically-separated stack of carbon foam (top and bottom layers) and aluminum foil plus RBF (center) after bonding.  The center piece is composed of 2 sheets of aluminum foil, 2 sheets of RBF, and another 2 sheets of aluminum foil all fused together by the RBF.  The rough texture of the aluminum foil is caused by hundreds of branching extrusions which locked into the carbon foam.}
  \label{fig:separated}
\end{figure}

Tensile strength tests were conducted with a pneumatic apparatus designed and fabricated in-house specifically for low ultimate tensile strength (UTS) samples (40 - 1500 kPa) \cite{thermalpaper} shown in Fig.~\ref{fig:tensile}.  Calibration was performed by adding known weights to the device and increasing the pressure until the piston assembly raised.  Resulting pressure versus sensor readout data are well-described by a simple linear function, shown in Fig.~\ref{fig:tensile}.
Samples were mounted using 3M 8810 pressure sensitive adhesive \cite{3Mtape} to attach them to large aluminum blocks, which were subsequently attached to steel hooks. These hooks were pulled with an increasing force by the tensile tester until the sample failed, breaking apart. The broken samples were observed for failure mode analysis, to determine how the bond differed from other samples.  An example is shown in Fig.~\ref{fig:separated}.

\section{Results and Discussion}
Ideal bonding conditions were achieved with 2 sheets of RBF and 2 sheets of aluminum foil with the mechanically locking clamp. Using less aluminum foil resulted in insufficient extruded aluminum to form a strong bond, while using less RBF resulted in insufficient heat for 2 aluminum foil sheets.  See Table~\ref{tab:results} for results for each sample configuration.  Error analysis shows UTS results are uncertain at a level of 10 kPa, independent of sample failure pressure.

Clamping method and clamping force affected tensile strength significantly.  However, this study did not vary clamping force and uniformity independently.  Thickness of the carbon foam did not affect results significantly for samples made with the locking clamp.  Failure mode analysis was performed by examining video recordings of tensile testing and examining samples to see which materials were removed from which surfaces. Failure tended to start in areas where the sample was not directly clamped and then proceed to rupture the rest of the bonded surface. For this reason, we believe uniformity of clamping force across the sample is key to a successful bond.  Bonding results are summarized as follows:

\begin{table}[!t]
\centering
\begin{tabular}{@{}|c|c|c|c|c|c|c|@{}}
\hline 
Sample & \# Sheets 25 $\mu{\rm m}$  & \# Sheets 60 $\mu{\rm m}$ & Carbon foam  & Clamp & UTS  \\
              & Al per side & RBF & thickness (mm) &  & (kPa) \\
\hline\hline
01 & 1 & 1 & 1  & 2x small clamp  & No bond \\\hline
02 & 1 & 1 & 4 &2x small clamp  & No bond \\\hline
03 & 1 & 2 & 1 & 2x small clamp  & No bond \\\hline
04 & 1 & 2 & 4 & 2x small clamp  & No bond \\\hline
05 & 2 & 1 & 1 & 2x small clamp  & No bond \\\hline
06 & 2 & 1 & 4 & 2x small clamp  & No bond \\\hline
07 & 2 & 2 & 4 & 2x small clamp & 45  \\\hline
08 & 2 & 2 & 1 & large clamp & 49 \\\hline
09 & 2 & 2 & 1 & locking clamp & > 124 \\\hline
10 & 2 & 2 & 1  & locking clamp & 140 \\\hline
11 & 2 & 2 & 4 & locking clamp & 147 \\
\hline

\end{tabular}
\caption{Ultimate tensile strength (UTS) results from various configurations of carbon foam-aluminum foil-RBF stacks.  Measurement uncertainties are described in the text.  Samples 01-06 fell apart when the clamps were removed.}
\label{tab:results}
\end{table}

\begin{figure}[!ht]
\centering
      \includegraphics[width=0.4\textwidth]{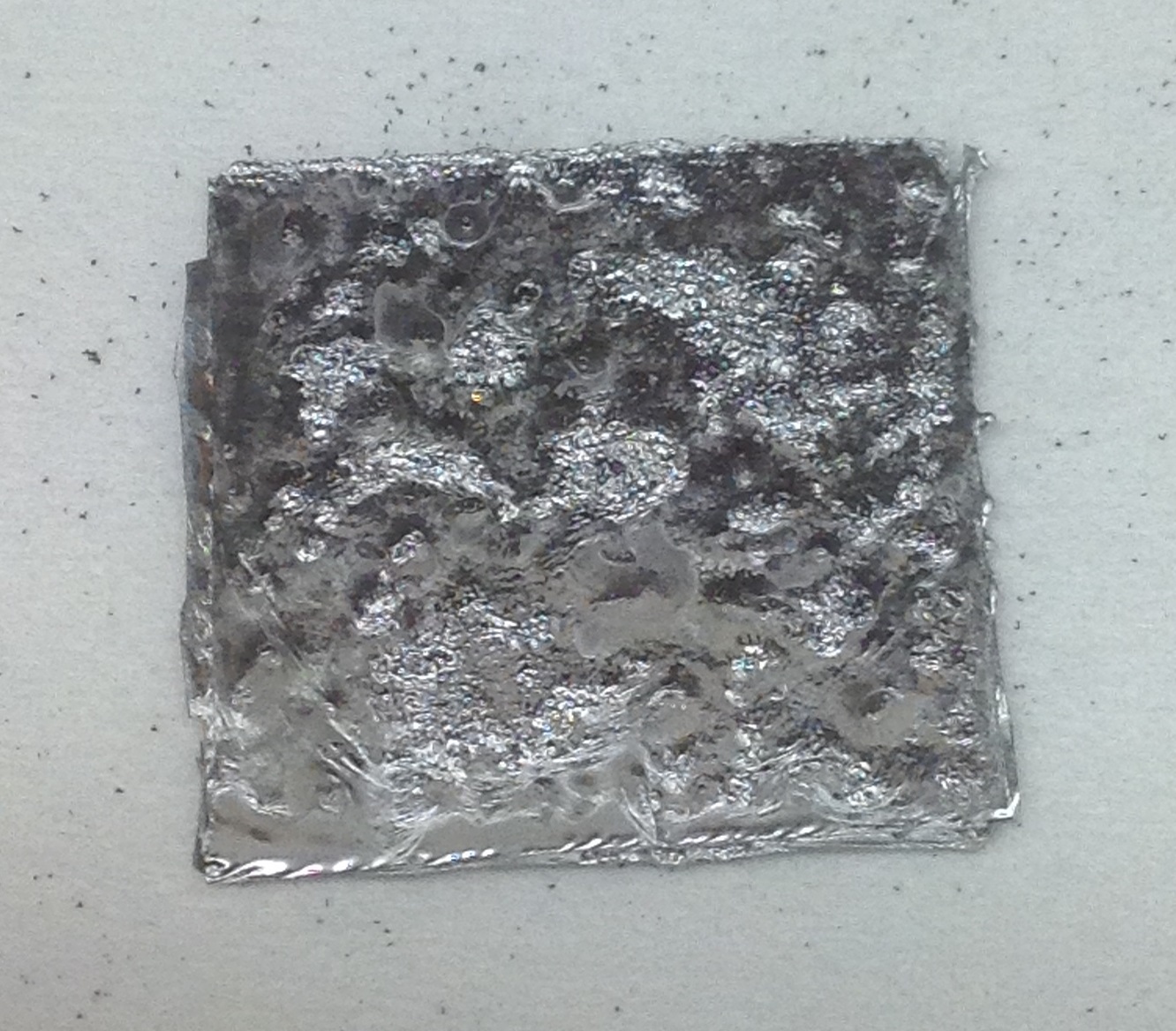}
    \includegraphics[width=0.4\textwidth]{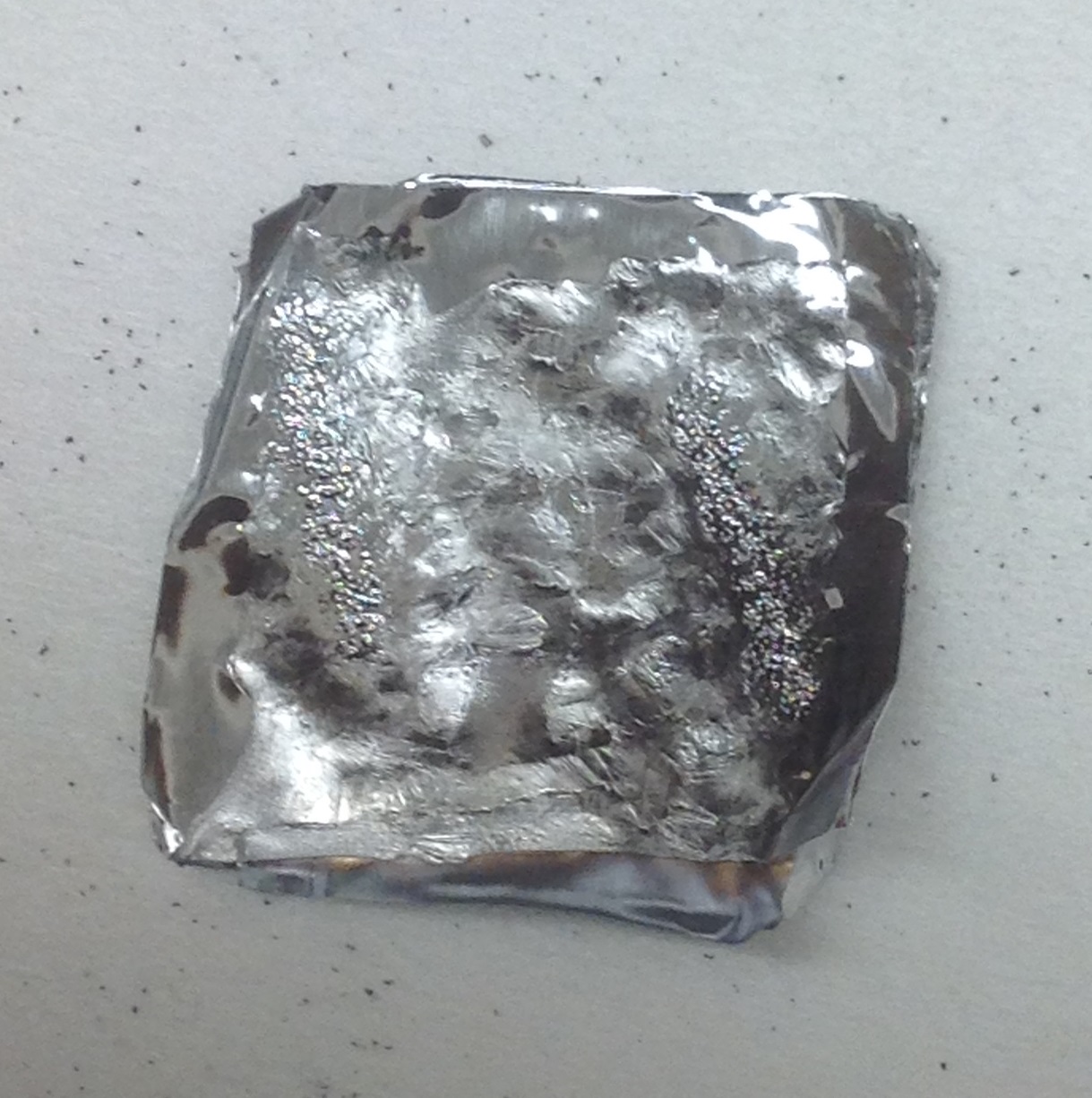}

  \caption{Left: sample with only 1 sheet of aluminum foil, which failed to extrude into foam.  Right: sample with only 1 sheet of RBF but 2 sheets of aluminum foil, for which aluminum did not melt completely.  Compare with Fig. 4, which demonstrates successful bonding.}
  \label{fig:problems}
\end{figure}

\begin{figure}[!ht]
\centering
    \includegraphics[width=0.8\textwidth]{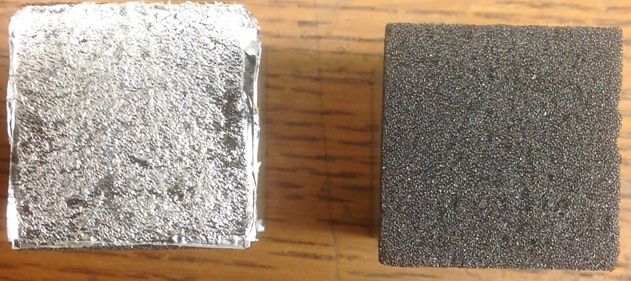}
  \caption{Sample after tensile testing.  Some holes are visible in the carbon foam where pieces of foam remained adhered to the aluminum.}
  \label{fig:aftertest}
\end{figure}

\begin{figure}[!ht]
\centering
      \includegraphics[width=0.4\textwidth]{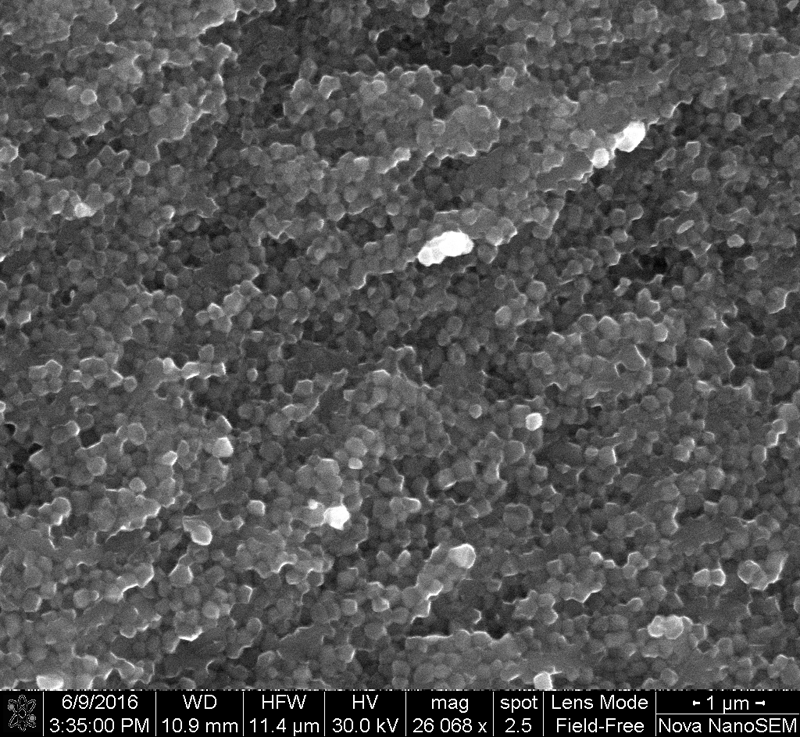}
    \includegraphics[width=0.4\textwidth]{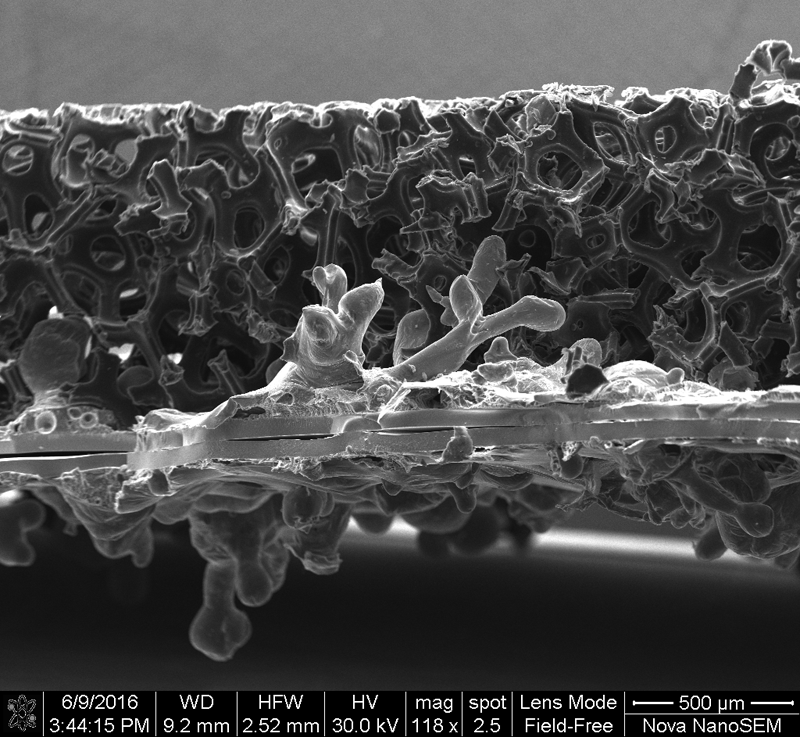}

  \caption{Left: SEM photo of RBF after reaction (fused ceramic).  Right: SEM photo of aluminum extrusions into carbon foam, demonstrating the "lock and key" structures responsible for successful bonding.  Structures reach up to 500 $\mu{\rm m}$ into the carbon foam matrix.}
  \label{fig:sem}
\end{figure}

\begin{itemize}

\item 
On samples with insufficient amount of aluminum foil (01-04), the critical "lock and key" structure could not be produced regardless of the other variables as there was not enough metal  (Fig.~\ref{fig:problems} Left).  Samples with 2 sheets of aluminum foil on each interface with only 1 sheet of RBF (05-06) did not produce the desired locking shape due to insufficient heat, i.e., the aluminum does not fully melt (Fig.~\ref{fig:problems} Right). 

\item
All samples with 2 sheets of RBF and 2 sheets of aluminum foil (07-11) created a mechanical bond (Figs.~\ref{fig:separated}, ~\ref{fig:aftertest}, and \ref{fig:sem}).  The strength of the bond increased with a strong, uniform clamping.  Clamps which did not accomplish this left the sample with weak points at the unclamped exposed portions, which did not bond as well.

\item
Sample 9 failed prematurely during testing due to failure of the adhesive between the aluminum testing block and the sample.  Upon inspection of the broken sample, it was found that the RBF interface also broke partially.  It can be concluded that when the adhesive on the testing block failed on one side, the pulling force produced a torque which lifted part of the sample at an angle, leading to premature failure of the bond.  Therefore, the actual tensile strength of the sample is likely greater than the measured value.

\item 
Our results give preliminary indication that carbon foam thickness does not affect bond strength significantly.  We hypothesized that thinner carbon foam structures may not provide enough thermal isolation or routes for air to escape during the reaction, however bond strength was consistent across thicknesses.
\end{itemize}

\section{Conclusions and Future Work}
We have investigated the use of reactive bonding film for carbon foam applications such as those for particle physics tracking detector mechanics.  This work shows that clamping mode is critical to obtaining a successful bond.  Preliminary results also indicate carbon foam thickness is not an important factor.  Successful bonds exhibited good strength due to the formation of metal extrusions in the foam.  

Though this work varied clamping strength and uniformity in tandem, future work may vary these separately to determine the precise clamping force and uniformity required for a successful bond.
Future work may also include the thermal and radiation tolerance characterization of  samples produced via this method.  We also intend to investigate this bonding technique for use with other materials such as carbon fiber and various metals.

\section{Acknowledgments}
RBF materials were provided as part of a research agreement with Indium Corporation.  This work at the University of California, Davis, was supported by U.S. Department of Energy grant DE-FG02-91ER40674 and by U.S. CMS R\&D funds via Fermilab.

\bibliographystyle{unsrt}

\end{document}